\documentclass[aip,apl,reprint]{revtex4-1}

\usepackage[utf8]{inputenc}
\usepackage{gensymb}
\usepackage{graphicx}
\usepackage[colorlinks=true,citecolor=blue,urlcolor=blue,linkcolor=black]{hyperref}
\pdfoutput=1

\begin{document}

\title{Sign change in the tunnel magnetoresistance of Fe\(_3\)O\(_4\)/MgO/Co-Fe-B magnetic tunnel junctions depending on the annealing temperature and the interface treatment} 

\author{L. Marnitz}
 \email{lmarnitz@physik.uni-bielefeld.de}
\author{K. Rott}
\author{S. Niehörster}
\author{C. Klewe}
\author{D. Meier}
\author{S. Fabretti}
\affiliation{Center for Spinelectronic Materials and Devices, Physics Department, Bielefeld University, Universitätsstraße 25, 33615 Bielefeld, Germany}

\author{M. Witziok}
\author{A. Krampf}
\author{O.~Kuschel}
\author{T. Schemme}
\author{K. Kuepper}
\author{J. Wollschläger}
\affiliation{Fachbereich Physik, Universität Osnabrück, Barbarastraße 7, 49069 Osnabrück, Germany}
\author{A. Thomas}
\author{G. Reiss}
\author{T. Kuschel}
\affiliation{Center for Spinelectronic Materials and Devices, Physics Department, Bielefeld University, Universitätsstraße 25, 33615 Bielefeld, Germany}

\date{\today}

\begin{abstract}
Magnetite (Fe\(_3\)O\(_4\)) is an eligible candidate for magnetic tunnel junctions (MTJs) since it shows a high spin polarization at the Fermi level as well as a high Curie temperature of 585\degree C. In this study, Fe\(_3\)O\(_4\)/MgO/Co-Fe-B MTJs were manufactured. A sign change in the TMR is observed after annealing the MTJs at temperatures between 200\degree C and 280\degree C. Our findings suggest an Mg interdiffusion from the MgO barrier into the Fe\(_3\)O\(_4\) as the reason for the change of the TMR. Additionally, different treatments of the magnetite interface (argon bombardment, annealing at 200\degree C in oxygen atmosphere) during the preparation of the MTJs have been studied regarding their effect on the performance of the MTJs. A maximum TMR of up to -12\% could be observed using both argon bombardment and annealing in oxygen atmosphere, despite exposing the magnetite surface to atmospheric conditions before the deposition of the MgO barrier.
\end{abstract}

\maketitle

\begin{figure}[t!]
\includegraphics[]{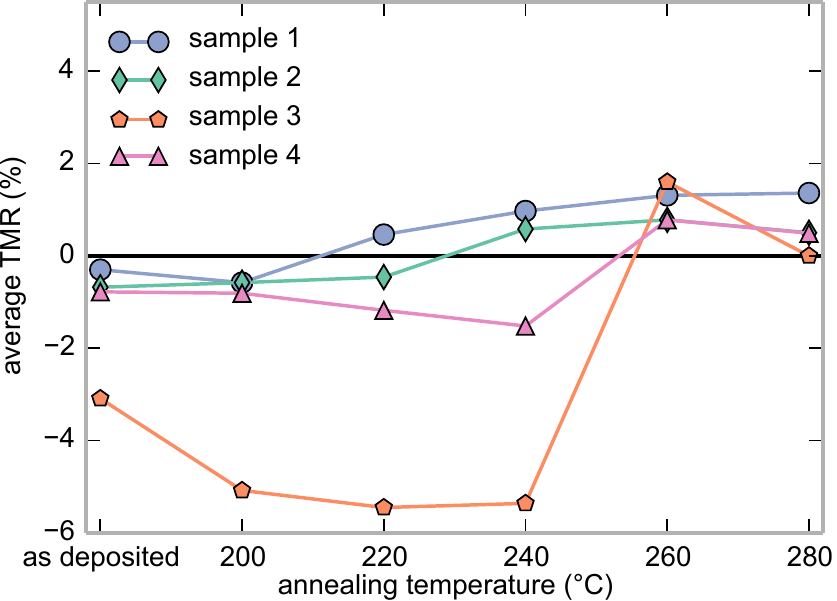}
\caption{Average TMR observed in MTJs depending on the annealing temperature. All samples start with a negative TMR and switch to a positive TMR after annealing at temperatures between 200\degree C and 260\degree C, independent of the magnetite treatment.}
\label{fig:average_TMR}
\end{figure}

The field of spintronics tries to create electronic devices which utilize the spin of the electron to store and process information \cite{Wolf2001}. A central device for this application is the magnetic tunnel junction, MTJ\cite{Yuasa2004}. It consists of two ferromagnetic conductors separated by a very thin insulating tunneling barrier. The resistance across this device depends on the relative orientation of the magnetization of the ferromagnetic electrodes, which leads to two different states: The magnetization of both electrodes can be parallel or antiparallel. This tunnel magnetoresistance (TMR) was first observed by Julliere et al.\cite{Julliere1975} in 1975. It is defined by \( \textrm{TMR}=\frac{R_{ap}-R_p}{R_p}\) where \(R_{ap}\) and \(R_p\) are the resistances for the antiparallel and parallel alignment of the magnetization of the electrodes. One application of this effect is using the corresponding high- and low-resistance states to store and read binary information.
\\
Magnetite (Fe\(_3\)O\(_4\)) shows a high spin polarization at the Fermi level\cite{Wang2013} as well as a high Curie temperature of 585\degree C \cite{Ziese2002}. It is thus an interesting material for MTJ electrodes as well as other spintronic and spin caloritronic applications. For example, magnetite has recently been studied regarding the spin Hall magnetoresistance \cite{Althammer2013,Ding2014} and the spin Seebeck and anomalous Nernst effect \cite{Ramos2013,Ramos2014,Wu2014,Wu2015}.
\\
However, MTJs with magnetite have not yet shown a large TMR. The largest reported TMR is \mbox{-26\%} at room temperature for Fe\(_3\)O\(_4\)/MgO/Al\(_2\)O\(_3\)/CoFe junctions while it also ranges up to +18\% in identical junctions\cite{Kado2008}. Table \ref{tab:TMR_Overview} shows an overview of reported TMR ratios from other studies using at least one Fe\(_3\)O\(_4\) electrode with different barrier and counter electrode materials. Especially the possibility of different signs of the TMR in junctions with the same stack suggest that the TMR is very sensitive to changes at the interfaces to the tunnel barrier \cite{DeTeresa:1999wx,Thomas:2005cj}. A negative TMR indicates different (i.e. one positive and one negative) signs of the spin polarization at the two electrode/barrier interfaces\cite{Klewe2013}.
\begin{table}[t!]\footnotesize
\renewcommand{\arraystretch}{1.2}
\begin{tabular}{|c|c|c|l|}
\hline 
Barrier & 2nd El. & TMR & Reference \\ 
\hline 
\hline
MgO/Al\(_2\)O\(_3\) & CoFe & -26\% & Kado\cite{Kado2008}, APL 2008 \\ 
&  & to +18\% &  \\ 
\hline 
Al\(_2\)O\(_3\) & CoFe & +14\% & Matsuda et al.\cite{Matsuda2002}, JJAP 2002 \\  
 &  &  & Aoshima, Wang \cite{Aoshima2003}, JAP 2003 \\ 
 &  &  & Yoon et al.\cite{Yoon2004}, JMMM 2005\\ 
\hline 
MgO & Fe\(_3\)O\(_4\) & +0.5\% & Li et al.\cite{Li1998}, APL 1998  \\ 
 &  &  & v. d. Zaag et al.\cite{vanderZaag2000}, JMMM 2000 \\ 
\hline 
Al\(_2\)O\(_3\) & Fe & -12\% & Nagahama et al.\cite{Nagahama2014}, APL 2014 \\ 
\hline 
Al\(_2\)O\(_3\) & Co & +13\% & Seneor et al.\cite{Seneor1999}, APL 1999 \\ 
\hline 
Al\(_2\)O\(_3\) & Co & +3\% & Bataille et al.\cite{Bataille2007}, JMMM 2007 \\ 
\hline 
AlO\(_x\) & Co & +20\% & Opel et al.\cite{Opel2011}, PSSA 2011 \\ 
\hline 
MgO & Co & -8\% & Greullet et al.\cite{Greullet2008}, APL 2008 \\ 
&  & to 0\%  & \\ 
\hline 
Al\(_2\)O\(_3\) & NiFe & -0.3\%  & Park et al.\cite{Park2005}, IEEE TM 2005 \\ 
&  & to +15\%  & \\ 
\hline 
Al\(_2\)O\(_3\) & Ni & +4\% & Reisinger et al.\cite{Reisinger2004}, Arxiv 2004 \\ 
\hline 
MgO & Ni & +0.5\% & Reisinger et al.\cite{Reisinger2004}, Arxiv 2004 \\ 
\hline 
CoCrl\(_2\)O\(_4\) & LSMO & -3\% & Hu, Suzuki\cite{Hu2002}, PRL 2002 \\ 
\hline 
\end{tabular} 
\caption{TMR ratios at room temperature of MTJs using one Fe\(_3\)O\(_4\) electrode and different barrier and counter electrode materials.}\label{tab:TMR_Overview}
\end{table}
\\
Gao et al. \cite{Gao1998} suggest that if magnetite is grown on MgO, Mg starts to diffuse into the magnetite at growth temperatures between 250\degree C and 350\degree C while Shaw et al.\cite{Shaw1997,Shaw2000} observe evidence of a starting interdiffusion of Mg from an MgO substrate into very thin (\(\sim\)10\(\,\)nm) magnetite layers at temperatures above 327\degree C. MgO is both used as a substrate for magnetite, because of a small lattice mismatch of 0.3\% \cite{Horng2004,Sterbinsky2007,Bertram2013} as well as a prominent barrier material for MTJs \cite{Parkin2004}, which in case of interdiffusion at the interface could lead to a diminished or altered TMR.
\\
In this study, Fe\(_3\)O\(_4\)/MgO/Co-Fe-B MTJs were prepared on MgO and the effect of this interdiffusion was studied for different annealing temperatures. Additionally, different treatments of the magnetite interface (argon
bombardment, annealing at 200\degree C in oxygen atmosphere) during the preparation of the MTJs have been studied regarding their effect on the performance of the MTJs.
\begin{table}[t!]\footnotesize
\renewcommand{\arraystretch}{1.2}
\begin{tabular}{|c|c|c|c|}
\hline 
Number & Ar bombardment & annealed in & Fe\(_3\)O\(_4\) thickness \\ 
\hline 
\hline
1 & no & UHV & 212nm \\ 
\hline 
2 & no & UHV & 40nm \\  
\hline 
3 & yes & O\(_2\) & 77nm \\ 
\hline 
4 & no & O\(_2\) & 108nm \\ 
\hline 
5 & yes & O\(_2\) & 22nm \\ 
\hline 
\end{tabular} 
\caption{This table shows the treatments applied in the preparation of each MTJ. Ar bombardment refers to bombarding the magnetite surface of the samples with Ar ions. Annealed in UHV/O\(_2\) refers the the first annealing step prior the the deposition of the MgO barrier and the subsequent layers. UHV means the sample was annealed for 2 hours in UHV at 200\degree C. O\(_2\) means the sample was annealed at 200\degree C for 2 hours in an oxygen atmosphere with a partial pressure of 10\(^{-4}\)\,mbar. Further annealing steps were done in UHV.}\label{tab:sample_treatments}
\end{table}
\\
Magnetite was grown on MgO(001) using reactive molecular beam epitaxy\cite{Bertram2012} (MBE). The subsequent layers were deposited using magnetron sputtering. The magnetite layers were exposed to atmospheric conditions during the transport from the MBE to the sputtering chamber. The corresponding thickness of the magnetite layer was measured using x-ray reflectivity employing a Philips X'Pert Pro diffractometer with a Cu K\(_{\alpha}\) source. The thickness for samples with the same treatment was varied to check for possible thickness dependencies, while for samples with different treatments the thickness was kept as similar as possible without being able to check the layer thickness in-situ during growth. Prior to the deposition of the MgO barrier and the other layers, the samples 1 and 2 were annealed at 200\degree C in UHV for 2 hours, while the samples 3, 4 and 5 were annealed at 200\degree C in an oxygen atmosphere with a partial pressure of 10\(^{-4}\)\,mbar. Additionally, the magnetite surface of the samples 3 and 5 were treated by argon bombardment before the annealing, to reduce surface contamination. The argon treatment was done with approximately \(4\, \mu \textrm{A/cm}^2\) at an energy of \(600\, \textrm{eV/ion}\) for 60 seconds. See table \ref{tab:sample_treatments} for an overview of the samples.
\begin{figure}[t!]
\includegraphics[]{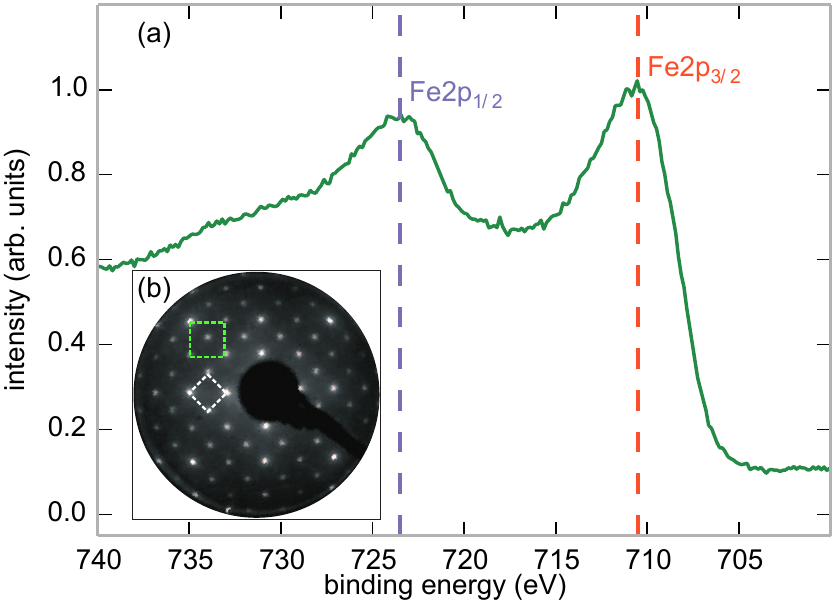}
\caption{\textbf{(a)} XP spectrum of a magnetite thin film. The binding energies of the Fe2p\(_{1/2}\) and Fe2p\(_{3/2}\) core levels correspond to values of Fe\(_3\)O\(_4\) (known from literature).  No charge transfer satelites are visible indicating the presence of mixed oxidation state of iron such as Fe\(_3\)O\(_4\) \cite{Fujii1999,Yamashita2008}. \textbf{(b)} LEED measurement of a magnetite thin film. The green square represents the reciprocal surface unit cell for magnetite while the small white square indicates the (\(\sqrt{2} \times \sqrt{2}\))R45\(^\circ\) superstructure which is reported for well-ordered magnetite \cite{Anderson1997,Korecki2002,Pentcheva2008}.}
\label{fig:xps_leed}
\end{figure}
\begin{figure}
\includegraphics[]{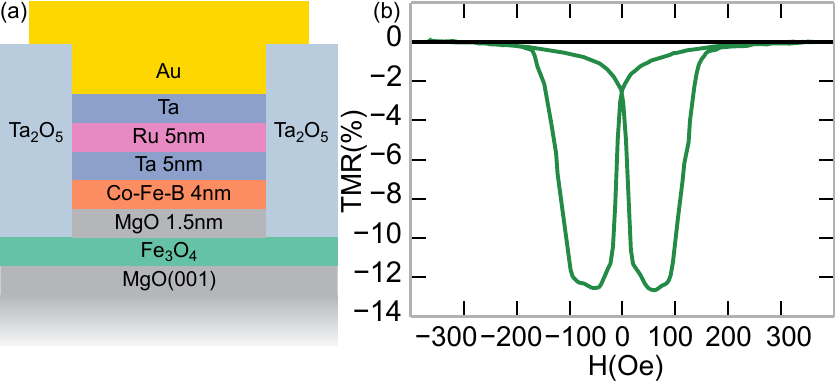}
\caption{\textbf{(a)} Schematic of the stacks (not to scale). The magnetite (Fe\(_3\)O\(_4\)) thickness was varied in each sample, see table \ref{tab:sample_treatments}. While the magnetite layer was deposited using MBE, all subsequent layers were grown by magnetron sputtering after a vacuum break.
\textbf{(b)} Maximal TMR measured on sample 5 after annealing at 230\degree C with a bias voltage of -200mV (average of 10 measurements).}
\label{fig:stackandmax}
\end{figure}
\begin{figure}
\includegraphics[]{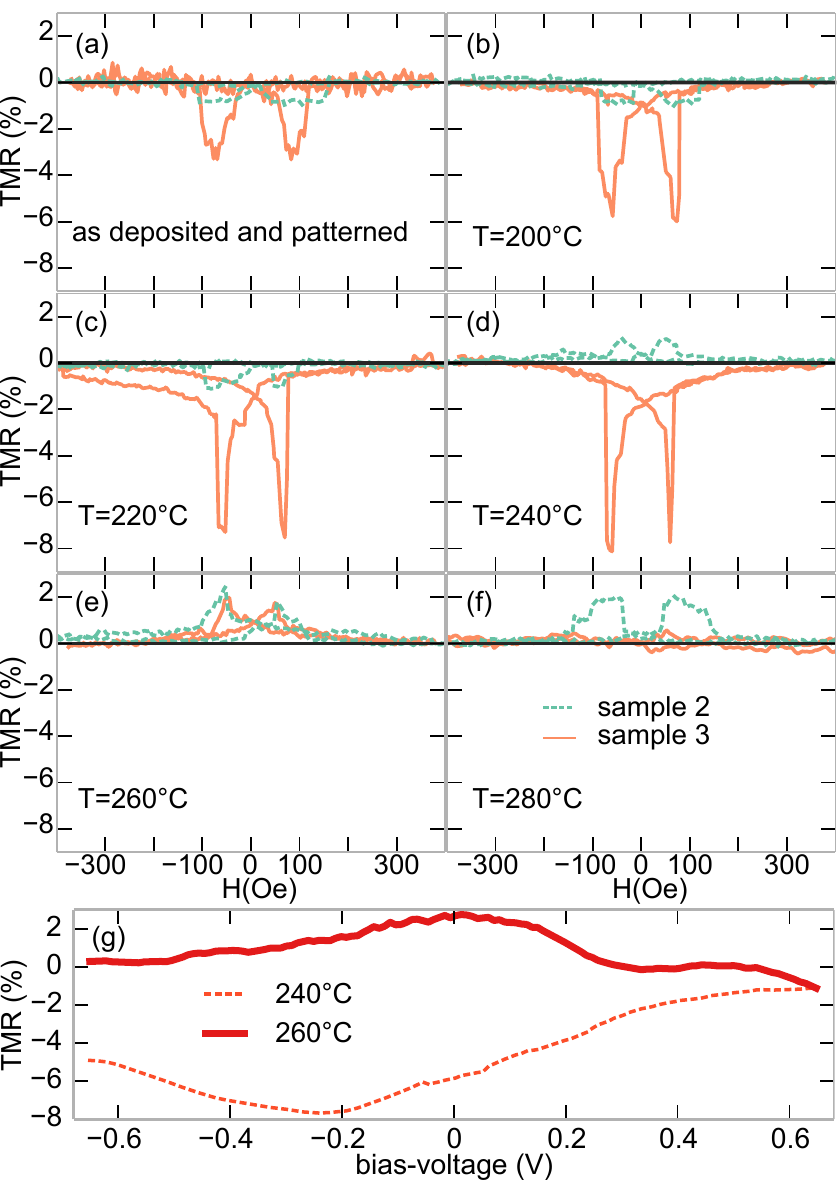}
\caption{TMR observed for the samples 2 and 3 \textbf{(a)} directly as deposited and patterned and after each annealing step from 200\degree C to 280\degree C in steps of 20\degree C at a bias voltage of 50mV for sample 2 and -200mV for 3 (\textbf{(b)-(f)} respectively). \textbf{(g)} shows the bias dependence of the TMR of sample 3 directly prior to and after the sign change. At lower annealing temperatures, the bias dependency looks very similar to the one shown for 240\degree C.}
\label{fig:annealing_steps}
\end{figure}
\\
Figure \ref{fig:xps_leed} shows the results of x-ray photoelectron spectroscopy (XPS) and low energy electron diffraction (LEED) measurements which were performed directly after the magnetite deposition in order to probe the quality of the magnetite surface near region. The binding energies of Fe2p\(_{1/2}\) and Fe2p\(_{3/2}\) deduced from XP spectra (cf. Fig. \ref{fig:xps_leed}(a)) show the values typical for Fe\(_3\)O\(_4\). Furthermore, no apparent charge transfer satellites can be observed in the spectra as it is well-known for Fe\(_3\)O\(_4\), in contrast to wüstite (FeO) and maghemite (Fe\(_2\)O\(_3\)) \cite{Fujii1999,Yamashita2008}.
\\
The LEED image (cf. Fig. \ref{fig:xps_leed}(b)) shows a typical diffraction pattern for a magnetite surface. The green square indicates the reciprocal surface unit cell of magnetite while the white square represents the (\(\sqrt{2} \times \sqrt{2}\))R45\(^\circ\) superstructure which is reported for well-ordered magnetite \cite{Anderson1997,Korecki2002,Pentcheva2008}.
\\
Figure \ref{fig:stackandmax}(a) shows a schematic of the MTJ stack used in this work. The area of the MTJs is elliptic with the major axis 1\,\(\mu\)m along the [110] axis of the magnetite, which is one of the magnetic easy axes \cite{Margulies1994}. The minor axis is 400\,nm long, while the top Au/Ta contact layer is 50\(\times\)50\,\(\mu\)m in size. Lithography was done using electron beam lithography and ion beam etching.
\\
The magnetoresistance measurements were performed on the samples \emph{as deposited} and after successively increased annealing temperatures starting at 200\degree C and increasing in steps of 20\degree C up to 280\degree C. Each time, at least 16 MTJs were measured. The bias voltage was set at 50\,mV for all samples not treated with argon bombardment. For all samples treated by argon bombardment, a bias voltage of -200\,mV showed the largest TMR effect while no sign change was found by varying the bias voltage. Hence, a bias voltage of -200\,mV was used for all these samples. For the other samples, there was only a very minor change in absolute TMR observed and the bias voltage was set at 50\,mV.
\begin{figure}[ht]
\includegraphics[]{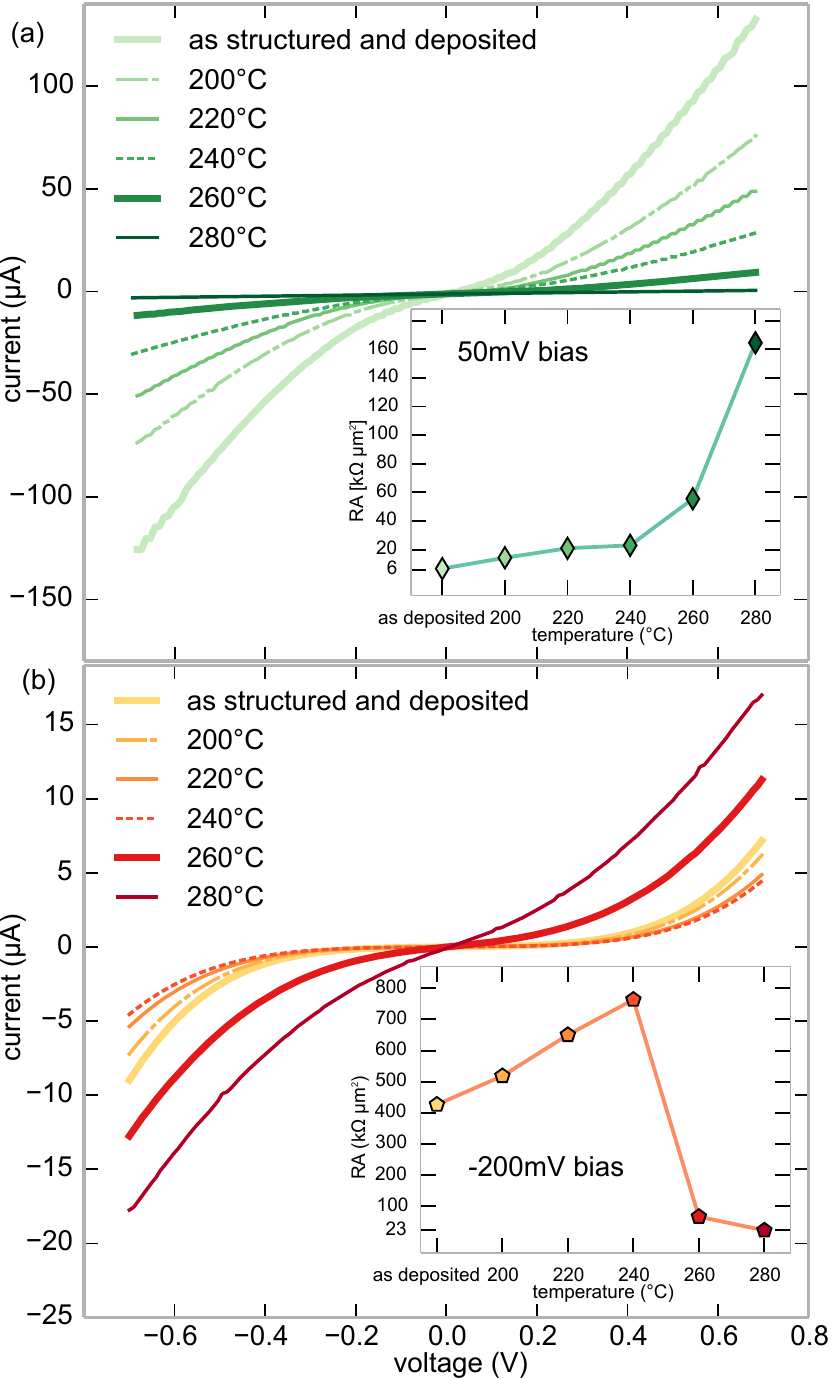}
\caption{I-V measurements for the samples \textbf{(a)} 2 and \textbf{(b)} 3, without applying an external magnetic field. \textbf{Insets:} The corresponding resistance-area products for parallel alignment at the bias voltages used for the measurements in figure \ref{fig:annealing_steps}. The resistance for sample 2 increases over the whole range of annealing temperatures, while for sample 3 the resistance drops after annealing at the same temperature at which the sign change in the TMR occurs.}
\label{fig:IVPlot}
\end{figure}
The largest TMR amplitude could be achieved in sample 5 with approximately -12\% TMR after annealing at 230\degree C (cf. Fig. \ref{fig:stackandmax}(b)). 
\\
Figure \ref{fig:annealing_steps} (a)-(f) shows a comparison of the samples 2 and 3 after each annealing step. In general, bombardment with argon ions greatly increases the TMR from values of about -1\% up to -12\%. Likely reasons for this are either a cleaner or a smoother magnetite surface (or a combination of both), leading to a better interface at the MgO barrier. A different interface between the magnetite electrode and the MgO barrier would also explain the differing behaviour to annealing between samples treated by argon bombardment and samples not treated. As shown in figure \ref{fig:annealing_steps} (a)-(f), both samples, with and without argon bombardment,  start with a negative TMR, switching to positive values at 240\degree C and 260\degree C respectively. However, after annealing at 280\degree C all MTJs of the sample treated by argon bombardment stop exhibiting any TMR. The bias dependency of the TMR of sample 3 is shown in figure \ref{fig:annealing_steps} (g) before and after the sign change.
\\
Figure \ref{fig:IVPlot} shows the corresponding I--V measurements, (a) and (b), for the same MTJs as in figure \ref{fig:annealing_steps}. The insets show the resulting resistance-area products (RA) for parallel magnetization of the electrodes at the bias voltages used for the TMR measurements in figure \ref{fig:annealing_steps}. It can be observed that the resistance of the sample treated by argon bombardment drops nearly by an order of magnitude after annealing at 260\degree C. At this temperature the sign of the TMR changes, too. For the sample not treated by argon bombardment, the resistance gradually increases with each annealing step, increasing rapidly after annealing at 240\degree C while no special feature can be seen after the TMR changes its sign. Additionally, the RA of sample 3 is up to 60 times as high as that of sample 2. This is attributed to the Fe\(_3\)O\(_4\)/MgO interface, since the growth conditions for both the magnetite and the MgO were identical for both samples. However, the difference in the treatment of the Fe\(_3\)O\(_4\) interface could lead to differing growth conditions of the MgO barrier.
\\
Figure \ref{fig:average_TMR} shows the characteristics of the MTJs after each annealing step. One can see that all samples start with a negative TMR regardless of the used treatments (annealing in O\(_2\)/UHV and argon bombardment) and then switch to a positive TMR after annealing at temperatures between 200\degree C and 260\degree C.
\\
This change is independent from the magnetite layer thickness, which suggests that Mg diffusion at the tunnel barrier might be responsible for the sign change instead of Mg diffusion from the substrate.
XPS studies on a magnetite film for annealing temperatures of up to 320\degree C under similar conditions than for the MTJs prepared (not shown here) showed no significant Mg2p signal which suggests that Mg diffusion from the substrate is not responsible for the sign change in the TMR but rather diffusion directly at the MgO barrier, which would explain the independence from the magnetite thickness.
\\
Annealing the magnetite samples in an oxygen atmosphere did not change the MTJs' behaviour significantly compared to annealing them in UHV. However, the sign change of sample 4 is observed at a higher annealing temperature than for samples 1 and 2. Again, the sample bombarded with argon ions shows a larger TMR compared to all other samples. All tested MTJs in the sample 3 stop working after annealing at 280\degree C, while samples not treated with argon bombardment still show a TMR. These findings again suggest that argon bombardment changes the interface at the tunnel barrier and that the sign change of the TMR occurs due to a modification of the interface at the tunnel barrier.
\\
In conclusion, Fe\(_3\)O\(_4\)/MgO/Co-Fe-B MTJs were shown to exhibit a TMR effect of up to -12\%, even though the magnetite layer was exposed to atmospheric conditions before the deposition of the MgO barrier. This was achieved by bombarding the magnetite surface with argon ions, likely leading to a cleaner interface. Annealing the magnetite samples in an oxygen atmosphere instead of UHV did not significantly improve the performance of the MTJs. It was also observed that after annealing the samples at temperatures between 200\degree C and 260\degree C, the sign of the TMR changed from negative to positive. A likely reason for this is the diffusion of Mg from the barrier into the magnetite, which changes its spin polarisation or modifies the barrier in a way that leads to different tunneling properties.

\begin{acknowledgments}
S.F., A.T. and T.K. were funded by the Ministry of Innovation, Science and Research (MIWF) of North Rhine-Westphalia through an independent researcher grant. We acknowledge support for the Article Processing Charge by the Deutsche Forschungsgemeinschaft and the Open Access Publication Fund of Bielefeld University. K.K. and G.R. acknowledge further support by the DFG (KU2321/2-1 and RE1052/32-1).
\end{acknowledgments}


\begin{thebibliography}{42}%
\bibitem {Wolf2001}{S.~A.~Wolf, D.~D.~Awschalom, R.~A.~Buhrman,J.~M.~Daughton, S.~von~Molnar, M.~L~Roukes, A.~Y.~Chtchelkanova, D.~M.~Treger, \href{http://dx.doi.org/10.1126/science.1065389}{Science~\textbf{294}, 1488 (2001)}}
\bibitem {Yuasa2004}{S.~Yuasa, {T.~Nagahama, A.~Fukushima, Y.~Suzuki, K.~Ando, \href{http://dx.doi.org/10.1038/nmat1257}{Nature Mater. \textbf{3}, 868 (2004)}
\bibitem {Julliere1975}{M.~Julliere, \href{http://dx.doi.org/10.1016/0375-9601(75)90174-7}{Phys. Lett. A}, \textbf{54}, 225 (1975)}}
\bibitem {Wang2013}{W.~Wang, J.~M.~Mariot, M.~C.~Richter, O.~Heckmann, W.~Ndiaye, P.~De~Padova, A.~Taleb-Ibrahimi, P.~Le~Fevre, F.~Bertran, F.~Bondino, E.~Magnano, J.~Krempasky, P.~Blaha, C.~Cacho, F.~Parmigiani, K.~Hricovini, \href{http://dx.doi.org/10.1103/PhysRevB.87.085118}{Phys. Rev. B \textbf{87}, 085118 (2013)}}
\bibitem {Ziese2002}{M.~Ziese, R.~Hohne, H.~C.~Semmelhack, H.~Reckentin, N.~H.~Hong, P.~Esquinazi, \href{http://dx.doi.org/10.1140/epjb/e2002-00245-3}{Eur. Phys. J. B \textbf{28}, 415 (2002)}}
\bibitem {Althammer2013}{M.~Althammer, S.~Meyer, H.~Nakayama, M.~Schreier, S.~Altmannshofer, M.~Weiler, H.~Huebl, S.~Geprägs, M.~Opel, R.~Gross, D.~Meier, C.~Klewe, T.~Kuschel, J.~M.~Schmalhorst, G.~Reiss, L.~Shen, A.~Gupta, Y.~T.~Chen, G.~E.~.W.~Bauer, E.~Saitoh, S.~T.~B.~Goennenwein, \href{http://dx.doi.org/10.1103/PhysRevB.87.224401}{Phys. Rev. B \textbf{87}, 224401 (2013)}}
\bibitem {Ding2014}{Z.~Ding, B.~L.~Chen, J.~H.~Liang, J.~Zhu, J.~X.~Li, Y.~Z.~Wu, \href{http://dx.doi.org/10.1103/PhysRevB.90.134424}{Phys. Rev. B \textbf{90}, 134424 (2014)}}
\bibitem {Ramos2013}{R.~Ramos, T.~Kikkawa, K.~Uchida, H.~Adachi, I.~Lucas, M.~H.~Aguirre, P.~Algarabel, L.~Morellón, S.~Maekawa, E.~Saitoh, M.~R.~Ibarra, \href{http://dx.doi.org/10.1063/1.4793486}{Appl. Phys. Lett. \textbf{102}, 072413 (2013)}}
\bibitem {Ramos2014}{R.~Ramos, M.~H.~Aguirre, A.~Anadón, J.~Blasco, I.~Lucas, K.~Uchida, P.~Algarabel, L.~Morellón, E.~Saitoh, M.~R.~Ibarra, \href{http://dx.doi.org/10.1103/PhysRevB.90.054422}{Phys. Rev. B \textbf{90}, 054422 (2014)}}
\bibitem {Wu2014}{S.~M.~Wu, J.~Hoffman, J.~E.~Pearson, A.~Bhattacharya, \href{http://dx.doi.org/10.1063/1.4895034}{Appl. Phys. Lett. \textbf{105}, 092409 (2014)}}
\bibitem {Wu2015}{S.~M.~Wu, F.~Y.~Fradin, J.~Hoffman, A.~Hoffmann, A.~Bhattacharya, \href{http://xxx.tau.ac.il/abs/1501.07599}{arXiv:1501.07599 [cond-mat.mes-hall] (2015)}}
\bibitem {Kado2008}{T.~Kado, \href{http://dx.doi.org/10.1063/1.2890852}{Appl. Phys. Lett. \textbf{92}, 092502 (2008)}}
\bibitem {Matsuda2002}{H.~Matsuda, M.~Takeuchi, H.~Adachi, M.~Hiramoto, N.~Matsukawa, A.~Odagawa, K.~Setsune, H.~Sakakima, \href{http://dx.doi.org/10.1143/JJAP.41.L387}{Jpn. J. Appl. Phys. \textbf{41}, L387 (2002)}} 
\bibitem {Aoshima2003}{K.~Aoshima, S.~X.~Wang, \href{http://dx.doi.org/10.1063/1.1558633}{J. Appl. Phys. \textbf{93}, 7954 (2003)}} 
\bibitem {Yoon2004}{K.~S.~Yoon, J.~H.~Koo, Y.~H.~Do, K.~W.~Kim, C.~O.~Kim, J.~P.~Hong, \href{http://dx.doi.org/10.1016/j.jmmm.2004.07.025}{J. Magn. Magn. Mater. \textbf{285}, 125 (2005)}} 
\bibitem {Li1998}{X.~Li, A.~Gupta, G.~Xiao, W.~Qian, V.~Dravid, \href{http://dx.doi.org/10.1063/1.122745}{Appl. Phys. Lett. \textbf{73}, 3282 (1998)}} 
\bibitem {vanderZaag2000}{P.~J.~van~der~Zaag, P.~J.~H.~Bloemen, J.~M.~Gaines, R.~M.~Wolf, P.~A.~A.~van~der~Heijden, R.~J.~M.~van~de~Veerdonk, W.~J.~M.~de~Jonge, \href{http://dx.doi.org/10.1016/S0304-8853(99)00751-9}{J. Magn. Magn. Mater. \textbf{211}, 301 (2000)}}  
\bibitem {Nagahama2014}{T.~Nagahama, Y.~Matsuda, K.~Tate, T.~Kawai, N.~Takahashi, S.~Hiratani, Y.~Watanabe, T.~Yanase, T.~Shimada, \href{http://dx.doi.org/10.1063/1.4894575}{Appl. Phys. Lett. \textbf{105}, 102410 (2014)}} 
\bibitem {Seneor1999}{P.~Seneor, A.~Fert, J.~L.~Maurice, F.~Montaigne, F.~Petroff, A.~Vaures, \href{http://dx.doi.org/10.1063/1.123246}{Appl. Phys. Lett. \textbf{74}, 4017 (1999)}} 
\bibitem {Bataille2007}{A.~M.~Bataille, R.~Mattana, P.~Seneor, A.~Tagliaferri, S.~Gota, K.~Bouzehouane, C.~Deranlot,  M.-J.~Guittet, J.-B.~Moussy, C.~de~Nadai, N.~B.~Brookes, F.~Petroff, M.~Gautier-Soyer, \href{http://dx.doi.org/10.1016/j.jmmm.2007.03.156}{J. Magn. Magn. Mater. \textbf{316}, E963 (2007)}} 
\bibitem {Opel2011}{M.~Opel, {S.~Gepraegs, E.~P.~Menzel, A.~Nielsen, D.~Reisinger, K.-W.~Nielsen, A.~Brandlmaier, F.~D.~Czeschka, M.~Althammer, M.~Weiler, S.~T.~B.~Goennenwein, J.~Simon, M.~Svete, W.~Yu, S.-M.~Huehne, W.~Mader, R.~Gross, \href{http://dx.doi.org/10.1002/pssa.201026403}{Phys. Status Solidi A \textbf{208}, 232 (2011)}}  
\bibitem {Greullet2008}{F.~Greullet, E.~Snoeck, C.~Tiusan, M.~Hehn, D.~Lacour, O.~Lenoble, C.~Magen, L.~Calmels, \href{http://dx.doi.org/10.1063/1.2841812}{Appl. Phys. Lett. \textbf{92}, 053508 (2008)}} 
\bibitem {Park2005}{C.~Park, J.~G.~Zhu, Y.~G.~Peng,  D.~E.~Laughlin, R.~M.~White, \href{http://dx.doi.org/10.1109/TMAG.2005.855294}{IEEE Trans. Magn. \textbf41}, 2691 (2005)}} 
\bibitem {Reisinger2004}{D.~Reisinger, P.~Majewski, M.~Opel, L.~Alff, R.~Gross \href{http://arxiv.org/abs/cond-mat/0407725v1}{arXiv:cond-mat/0407725v1} (2004)}}
\bibitem {Hu2002}{G.~Hu, Y.~Suzuki, \href{http://dx.doi.org/10.1103/PhysRevLett.89.276601}{Phys. Rev. Lett. \textbf{89}, 276601 (2002)}} 
\bibitem {DeTeresa:1999wx}{J.~De~Teresa, A.~Barth{\'e}l{\'e}my, A.~Fert, J.~Contour, \href{http://dx.doi.org/10.1126/science.286.5439.507}{Science \textbf{286}, 507 (1999)}}
\bibitem {Thomas:2005cj}{A.~Thomas, J.~S.~Moodera, B.~Satpati, \href{http://dx.doi.org/10.1063/1.1850400}{J. Appl. Phys. \textbf{97}, 10C908 (2005)}}
\bibitem {Klewe2013}{C.~Klewe, M.~Meinert, J.~Schmalhorst, G.~Reiss, \href{http://stacks.iop.org/0953-8984/25/i=7/a=076001}{J. Phys. Cond. Matter \textbf{25}, 076001 (2013)}}
\bibitem {Gao1998}{Y.~Gao, Y.~J.~Kim, S.~A.~Chambers, \href{http://dx.doi.org/10.1557/JMR.1998.0281}{J. Mater. Res. \textbf{13}, 2003 (1998)}} 
\bibitem {Shaw1997}{K.~A.~Shaw, E.~Lochner, D.~M.~Lind, J.~F.~Anderson, M.~Kuhn, U.~Diebold, \href{http://dx.doi.org/10.1063/1.365161}{J. Appl. Phys. \textbf{81}, 5176 (1997)}} 
\bibitem {Shaw2000}{K.~A.~Shaw, E.~Lochner, D.~M.~Lind, \href{http://dx.doi.org/10.1063/1.372084}{J. Appl. Phys. \textbf{87}, 1727 (2000)}} 
\bibitem {Horng2004}{L.~Horng, G.~Chern, M.~C.~Chen, P.~C.~Kang, D.~S.~Lee, \href{http://dx.doi.org/10.1016/j.jmmm.2003.09.005}{J. Magn. Magn. Mater. \textbf{270}, 389 (2004)}} 
\bibitem {Sterbinsky2007}{G.~E.~Sterbinsky, J.~Cheng, P.~T.~Chiu, B.~W.~Wessels, D.~J.~Keavney, \href{http://dx.doi.org/10.1116/1.2757185}{J. Vac. Sci. Technol., B \textbf{25}, 1389 (2007)}} 
\bibitem {Bertram2013}{F.~Bertram, C.~Deiter, T.~Schemme, S.~Jentsch, J.~Wollschläger, \href{http://dx.doi.org/10.1063/1.4803894}{J. Appl. Phys. \textbf{113}, 184103 (2013)}} 
\bibitem {Parkin2004}{S.~S.~P.~Parkin, C.~Kaiser, A.~Panchula, P.~M.~Rice, B.~Hughes, M.~Samant, S.~H.~Yang, \href{http://dx.doi.org/10.1038/nmat1256}{Nature Mater. \textbf{3}, 862 (2004)}} 
\bibitem {Bertram2012}{F.~Bertram, C.~Deiter, O.~Hoefert, T.~Schemme, F.~Timmer, M.~Suendorf, B.~Zimmermann, J.~Wollschläger, \href{http://dx.doi.org/10.1088/0022-3727/45/39/395302}{J. Phys. D: Appl. Phys. \textbf{45}, 395302 (2012)}} 
\bibitem {Fujii1999}{T.~Fujii, F.~M.~F.~de~Groot, G.~A.~Sawatzky, F.~C.~Voogt, T.~Hibma, K.~Okada, \href{http://dx.doi.org/10.1103/PhysRevB.59.3195}{Phys. Rev. B \textbf{59}, 3195 (1999)}} 
\bibitem {Yamashita2008}{T.~Yamashita, P.~Hayes, \href{http://dx.doi.org/10.1016/j.apsusc.2007.09.063}{Appl. Surf. Sci. \textbf{254}, 2441 (2008)}} 
\bibitem {Anderson1997}{J.~F.~Anderson, M.~Kuhn, U.~Diebold, K.~Shaw, P.~Stoyanov, D.~Lind, \href{http://dx.doi.org/10.1103/PhysRevB.56.9902}{Phys. Rev. B \textbf{56}, 9902 (1997)}} 
\bibitem {Korecki2002}{J.~Korecki, B.~Handke, N.~Spiridis, T.~Slezak, I.~Flis-Kabulska, J.~Haber, \href{http://dx.doi.org/10.1016/S0040-6090(02)00306-1}{Thin Solid Films \textbf{412}, 14 (2002)}} 
\bibitem {Pentcheva2008}{R.~Pentcheva, W.~Moritz, J.~Rundgren, S.~Frank, D.~Schrupp, M.~Scheffler, \href{http://dx.doi.org/10.1016/j.susc.2008.01.006}{Surf. Sci. \textbf{602}, 1299 (2008)}} 
\bibitem {Margulies1994}{D.~T.~Margulies, F.~T.~Parker, A.~E.~Berkowitz, \href{http://dx.doi.org/10.1063/1.355472}{J. Appl. Phys. \textbf{75}, 6097 (1994)}} 
\end{thebibliography}
\end{document}